\documentclass[prb,nofootinbib,superscriptaddress,twocolumn,showpacs]{revtex4-1}

\usepackage{bm}
\usepackage{amsfonts}
\usepackage[usenames,dvipsnames]{color}
\usepackage{graphicx}
\usepackage{amssymb}
\usepackage{amsmath}
\usepackage{bbm,dsfont}
\usepackage{amsthm}
\usepackage{hyperref}
\hypersetup{
        colorlinks=true,
}

\usepackage{times}
\usepackage{braket}

\usepackage[english]{babel}
\usepackage{blindtext}

\newcommand{\vect}[1]{\mathbf{#1}}

\newcommand{\nn}{\nonumber}

\newcommand{\Ba}{Ba$_3$IrTi$_2$O$_9$}

\let\oldmarginpar\marginpar
\renewcommand\marginpar[1]{\-\oldmarginpar[\raggedleft\tiny\color{red} #1]%
{\raggedright\tiny #1}}

\newcommand{\citationtitle}[1]{{\color{Gray}\small #1}}

\graphicspath{{Figures/}}

\begin{document}

\title{Spin-orbit physics of j=1/2 Mott insulators on the triangular lattice}
\date{\today}

\author{Michael Becker}
\author{Maria Hermanns}
\affiliation{Institute for Theoretical Physics, Cologne University, Z\"ulpicher Stra\ss e 77, 50937 Cologne, Germany}
\author{Bela Bauer}
\affiliation{Station Q, Microsoft Research, Santa Barbara, CA 93106-6105, USA}
\author{Markus Garst}
\author{Simon Trebst}
\affiliation{Institute for Theoretical Physics, Cologne University, Z\"ulpicher Stra\ss e 77, 50937 Cologne, Germany}


\begin{abstract}
The physics of spin-orbital entanglement in effective $j=1/2$ Mott insulators, which have been experimentally observed for various 5d transition metal oxides, has sparked an interest in Heisenberg-Kitaev (HK) models thought to capture their essential microscopic interactions. 
Here we argue that the recently synthesized Ba$_3$IrTi$_2$O$_9$ is a prime candidate for a microscopic realization of the {\em triangular} HK model -- a conceptually interesting model for its interplay of geometric and exchange frustration. 
We establish that an infinitesimal Kitaev exchange destabilizes the 120$^\circ$ order of the quantum Heisenberg model. This results in the formation of an extended $\mathbb{Z}_2$-vortex crystal phase in the parameter regime most likely relevant to the real material, which can be experimentally identified with spherical neutron polarimetry.
Moreover, using a combination of analytical and numerical techniques
we map out the entire phase diagram of the model, which further includes various ordered phases as well as  an extended nematic phase around the antiferromagnetic Kitaev point.
\end{abstract}

\maketitle


\section{Introduction}
The physics of transition metal oxides with partially filled $5d$ shells is governed by a largely accidental balance of electronic correlations, spin-orbit entanglement, and crystal field effects, with all three components coming up roughly equal in strength. With different materials exhibiting slight tilts towards one of the three effects a remarkably broad variety of quantum states has recently been suggested, which includes exotic states such as Weyl semi-metals, axion insulators, or topological Mott insulators \cite{witczak}.
A particularly intriguing scenario is the formation of Mott insulators in which the local moments are spin-orbit entangled Kramers doublets.
An example are the $j=\frac{1}{2}$ Mott insulators observed for various Iridates \cite{kim08,kim09,singh2012}. The Iridium valence in the latter typically is Ir$^{4+}$ corresponding to a $5d^5$ electronic configuration. With the crystal field of the octahedral IrO$_6$ oxygen cage splitting off the two e$_g$ levels, this puts 5 electrons with an effective $s=\frac{1}{2}$ magnetic moment into the t$_{2g}$ orbitals, which entangled by strong spin-orbit coupling leaves the system with a fully filled $j=\frac{3}{2}$ band and a half-filled $j=\frac{1}{2}$ band \cite{kim08,kim09,pesin10}. The smaller bandwidth of the latter then allows for the opening of a Mott gap even for the relatively moderate electronic correlations of the $5d$ compounds.
Interest in such $j=\frac{1}{2}$ Mott insulators has been sparked by the theoretical observation \cite{Khaliullin2005,jackeli2009,chaloupka10} that the  microscopic interaction between their spin-orbit entangled local moments not only includes an isotropic Heisenberg exchange but also highly anisotropic interactions whose easy axis depends on the spatial orientation of the exchange path tracing back to the orbital contribution of the moments \cite{kugel1982}. In a hexagonal lattice geometry, as it is found for the layered Na$_2$IrO$_3$ and $\alpha$-Li$_2$IrO$_3$ compounds, these anisotropic interactions provide an implementation of the celebrated Kitaev model \cite{kitaev2006} known for its spin liquid ground states. A trove of experimental data \cite{exp}, ab initio calculations \cite{abinitio}, and model simulations \cite{models} for these hexagonal systems has spurred an ongoing discourse illuminating the actual spin-orbital ordering mechanism in these materials.

Much recent activity \cite{Kimchi2014,beyond} has been targeted towards the physics of $j=1/2$ Mott insulators for lattice geometries beyond the hexagonal lattice, triggered mainly by the synthesis of novel Iridate compounds, which includes e.g. the sister compounds $\beta$-Li$_2$IrO$_3$ \cite{beta} and $\gamma$-Li$_2$IrO$_3$ \cite{gamma} that form three-dimensional Ir lattice structures.
In this manuscript, we turn to the recently synthesized  Iridate \Ba\ \cite{dey2012} and argue that it realizes a 
Heisenberg-Kitaev (HK) model on a {\em triangular} lattice\cite{vesta}.
This model is of deep conceptual interest as it exhibits a subtle interplay of the two elementary sources of frustration -- geometric frustration arising from its non-bipartite lattice structure as well as exchange frustration arising from the Kitaev couplings. 
The ground states of the classical HK model on the triangular lattice have been already addressed by Rousochatzakis {\it et al.} \cite{rous2012}. With the help of Monte Carlo simulations these authors demonstrated that a small but finite Kitaev exchange in addition to an antiferromagnetic Heisenberg interaction stabilizes a $\mathbb{Z}_2$-vortex crystal. The $\mathbb{Z}_2$-vortices can be viewed as defects of the SO(3) order parameter associated with the 120$^\circ$ ordering of the antiferromagnetic Heisenberg model.
It is important to note that this physics plays out near the Heisenberg limit of the HK model -- the relevant microscopic parameter regime for all Iridates synthesized so far. As such \Ba\ is a prime candidate to observe this exotic phase. 

After a discussion of the material aspect of \Ba\ and a motivation of the HK model in Sec. \ref{sec:Material}, we examine in Sec. \ref{sec:VortexCrystal} the formation of a $\mathbb{Z}_2$-vortex crystal from the analytical perspective of an expanded Luttinger-Tisza approximation and quantitatively describe its experimental signatures in polarized neutron scattering experiments. In Sec. \ref{sec:PhaseDiagram} we address the full phase diagram of the HK model and discuss the various phases with the help of analytical as well as numerical methods. 
Finally, in Sec. \ref{sec:Conclusions} we close with a summary.


\section{Material physics of $\mbox{\Ba}$}
\label{sec:Material}

\begin{figure*}[t]
	\includegraphics[width=\linewidth]{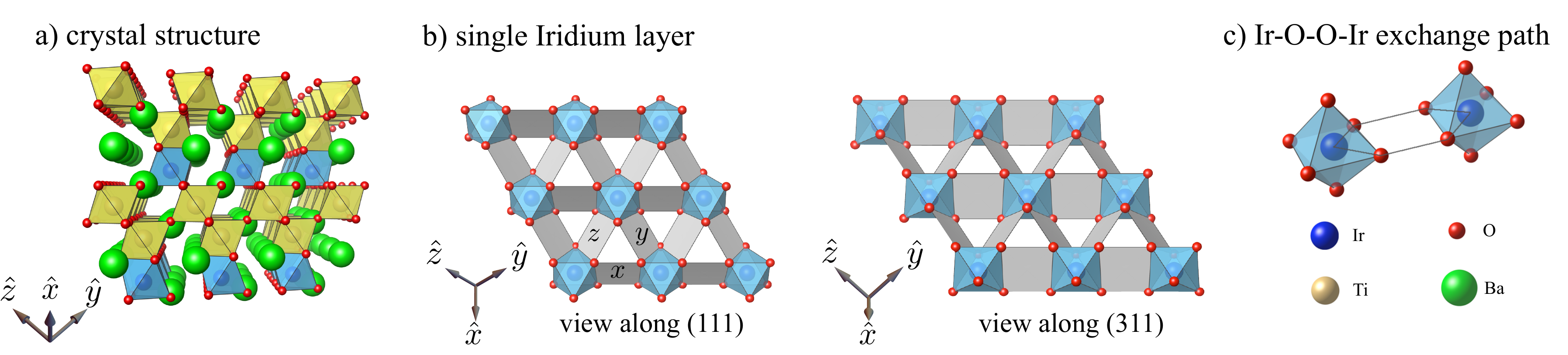}
	\caption{(Color online) 
	              a) Crystal structure of Ba$_3$IrTi$_2$O$_9$. 
	              b) View of single Iridium layers from two different perspectives. 
	              	  Within the plane the $x$, $y$, and $z$ exchange paths are indicated by the grey planes. The planes labeled by $x$ ($y$, $z$) are normal to the coordinate axis $\hat x$ ($\hat y$, $\hat z$). 
	              c) The exchange between the Iridium moments (blue) is mediated by two coplanar exchange paths.}
        \label{fig:iridate}
\end{figure*}

\Ba\ forms layers of Ir$^{4+}$ ions in a triangular geometry, which are separated from each other by two layers of Ti$^{4+}$ ions. 
An important characteristic of the Ir layer geometry illustrated in Figs.~\ref{fig:iridate}~a) and b) is that it exhibits the two necessary ingredients for Kitaev-type exchange couplings.
First, every pair of Iridium ions is coupled via {\em two} separate exchange paths as indicated in Fig.~\ref{fig:iridate}~c) leading to a destructive interference and subsequent suppression of the isotropic Heisenberg exchange \cite{Khaliullin2005,jackeli2009,chaloupka10}. 
In comparison to the tricoordinated Iridates (Na,Li)$_2$IrO$_3$, which exhibit Ir-O-Ir exchange paths, the triangular \Ba\  exhibits somewhat longer Ir-O-O-Ir exchange paths resulting in an overall lessening of the magnetic exchange strength.
Second, the three principal bond directions of the triangular lattice structure cut through three {\em different} edges of the IrO$_6$ oxygen cages resulting in a distinct locking of the exchange easy axis along the three directions \cite{Khaliullin2005,jackeli2009,chaloupka10} as illustrated Fig.~\ref{fig:iridate} a) and ultimately giving rise to the three components of the Kitaev exchange. Note that the Ir layer is normal to the (111) direction, hence the three directions are all equivalent. 
The description of the  microscopic physics is thus given in terms of a Heisenberg-Kitaev (HK) Hamiltonian 
\begin{equation} \label{Ham}
\mathcal{H}_{\rm HK}=J_H\sum_{\braket{ij}}\vect{S}_i\cdot\vect{S}_j + J_K\sum_{\gamma\parallel\braket{ij}} S^\gamma_i S^\gamma_j,
\end{equation}
where $\vect{S}_i$ is a spin-operator located on site $i$ of the triangular lattice spanned by the lattice vectors $\vect a_x = (1,0)^T$, $\vect a_y = (-1/2,\sqrt{3}/2)^T$, and $\vect a_z = - \vect a _x - \vect a_y$, see Fig.~\ref{fig:lattice}~a). Here and in the following, we measure lengths in units of the lattice constant $a$. The first term is the standard Heisenberg coupling, $J_H$, that describes an SU(2) invariant interaction between the spin-orbit entangled $j=1/2$ moments on nearest-neighbor lattice sites. The Kitaev interaction, $J_K$, on the other hand,  explicitly breaks spin-rotation invariance and acts only between single components, $S^\gamma$, of adjacent spins. The precise component   depends on the link between the lattice sites, see Fig.~\ref{fig:lattice}~a); for our particular choice here, the $\gamma$-components of spins interact via $J_K$ if sites are connected by a lattice vector $\vect a_\gamma$ with $\gamma=x,y,z$.


\section{120$^\circ$ order and $\mathbb{Z}_2$-vortex crystal}
\label{sec:VortexCrystal}

We will start our discussion of the ground states of Hamiltonian  \eqref{Ham} by first elucidating the magnetic structure around the antiferromagnetic Heisenberg point, 
where an extended $\mathbb{Z}_2$ vortex crystal phase is found in agreement with Ref.~\cite{rous2012}.
The ground state of the antiferromagnetic Heisenberg  Hamiltonian on the triangular lattice, which corresponds to couplings $J_H>0$ and $J_K = 0$ for Hamiltonian \eqref{Ham},  is characterized by a 120$^\circ$ ordering of spins \cite{120NeelOrder}.  At the classical level this ordering is captured by a spin orientation $\vect{S}_i = S \vect{\hat\Omega}(\vect r_i)$ with the unit vector $\vect{\hat \Omega}_{120^\circ}(\vect r) = \vect{e}_1 \cos \left(\vect Q \cdot\vect r\right) + \vect{e}_2 \sin\left(\vect Q\cdot \vect r\right)$ where the commensurate wavevector $\vect Q$ connects the center with a corner of the Brillouin zone, 
$\vect Q = \frac{4\pi}{3} (1,0)$. The orthonormal frame $\vect e_i$ with $i=1,2,3$ and $\vect e_3 = \vect e_1 \times \vect e_2$ constitutes an SO(3) order parameter. The energy per site for this classical state  is given by 
\begin{align} \label{120Energy}
\varepsilon_{120^\circ}
=  - S^2 \frac{1}{2} \Big(3 J_H + J_K \Big).
\end{align}

Crucially, the 120$^\circ$ ordering possesses $\mathbb{Z}_2$ vortices \cite{kawamura84} as topologically stable point defects, which can be understood by considering the first homotopy group of its order parameter $\Pi_1{\rm (SO(3))} = \mathbb{Z}_2$.


\subsection{Kitaev interaction destabilizes 120$^\circ$ ordering}


For any finite $J_K$ the 120$^\circ$ state becomes immediately unstable with respect to fluctuations, which we demonstrate in the following. We parametrize the fluctuations with the help of two real fields $\bm \pi(\vect r) = (\pi_1(\vect r),\pi_2(\vect r))^T$,
\begin{align}
\vect{\hat \Omega}(\vect r) &= \vect{\hat \Omega}_{120^\circ}(\vect r) \sqrt{1 - \left(\bm{\pi}(\vect r)\right)^2} 
\\\nn&
+ \pi_1(\vect r) \left(- \vect{e}_1 \sin\left(\vect Q\cdot \vect r\right) + \vect{e}_2 \cos\left( \vect Q \cdot\vect r \right)\right) + \pi_2(\vect r) \vect{e}_3,
\end{align}
so that $\vect{\hat \Omega}^2(\vect r) = 1$ is mantained. Plugging this Ansatz in the Hamiltonian and expanding up to second order in the fluctuation fields one obtains for the energy $\mathcal{E} = N \varepsilon_{120^\circ} + \mathcal{E}^{(2)}$ with $N$ denoting the number of lattice sites. The fluctuation part reads
\begin{align}
\nn
&\mathcal{E}^{(2)} = 
-\varepsilon_{120^\circ} \sum_i \left(\bm{\pi}(\vect r_i)\right)^2
-\frac{J_H S^2}{2} \sum_{\braket{ij}} (\pi_{1i} \pi_{1j} - 2 \pi_{2i} \pi_{2j}) 
\\\nn&
+ J_K S^2 \sum_{\gamma\parallel\braket{ij}}  \Big[ 
e^\gamma_3  e^\gamma_3 \pi_{2i} \pi_{2j} + \Big( e^\gamma_1 e^\gamma_1 \cos(\vect Q\vect r_i) \cos(\vect Q\vect r_j) 
\\\nn& 
+ e^\gamma_2  e^\gamma_2 \sin(\vect Q\vect r_i) \sin(\vect Q\vect r_j) 
- e^\gamma_1  e^\gamma_2 \sin(\vect Q(\vect r_i + \vect r_j))
\Big) \pi_{1i} \pi_{1j} 
\\&
+ \Big(
\Bigl[- e^\gamma_1 \sin(\vect Q\vect r_i)+  e^\gamma_2 \cos(\vect Q\vect r_i)  \Bigr]  e^\gamma_3 \pi_{1i} \pi_{2j}  
+ (i \leftrightarrow j) \Big)
\Big],
\end{align}
with the abbreviation $\pi_{ai} = \pi_{a}(\vect r_i)$ for $a=1,2$. The fluctuation eigenmodes are determined with the help of the Fourier transform $\pi_{a}(\vect r_i) = \frac{1}{\sqrt{N}}\sum_{{\bf k} \in {\rm 1. BZ}} e^{i {\mathbf{ k r}_i}} \pi_a({\bf k})$. In the absence of the Kitaev interaction, $J_K = 0$, one obtains
\begin{align}
\mathcal{E}^{(2)}\Big|_{J_K = 0} &= \frac{J_H S^2}{2} 
\sum_{{{\bf k} \in {\rm 1. BZ}}\atop{\gamma = x,y,z}}
\Big[ (1- \cos(\vect k \vect a_\gamma)) \pi^*_{1}({\bf k})\pi_{1}({\bf k})
\nn\\&+ (1+ 2 \cos(\vect k\vect a_\gamma)) \pi^*_{2}({\bf k}) \pi_{2}({\bf k})\Big]
\end{align}
with $ \pi^*_a({\bf k}) = \pi_a(-{\bf k})$.
Whereas the $\pi_1$ mode becomes soft at the center of the Brillouin zone, i.e., at $\vect k = 0$, the energy of the $\pi_2$ mode vanishes at its edge, e.g.\ for momenta $\vect k = \pm \vect Q$. The zero modes $\pi_1(\vect k=0)$ and $\pi_2(\pm\vect Q)$ thus identify three Goldstone modes that correspond to a long-wavelength rotation and tilting of the local orthogonal frame, respectively. In particular, the energy dispersion of the tilting mode $\varepsilon^{\rm tilt}_{\bf k}|_{J_K=0} = J_H S^2\sum_{\gamma = 1,2,3} (1+ 2 \cos(\vect k  \cdot\vect a_\gamma))$ close to momentum $\vect Q$ possesses the form $\varepsilon^{\rm tilt}_{\bf Q+k}|_{J_K=0} \approx J_H S^2 \tfrac{3}{8} {\bf k}^2$.

\begin{figure}
\includegraphics[width=.94\linewidth]{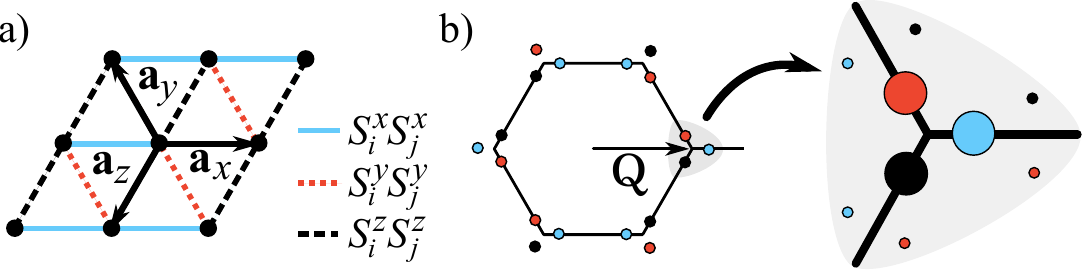}
\caption{(Color online) 
a) The triangular lattice with the three lattice vectors $\vect{a}_\gamma$. Solid, dashed and dotted bonds carry the three distinct Kitaev interactions, see text. 
b) First Brillouin zone of the triangular lattice. The position and size of the coloured dots indicate the position and weight of Bragg peaks, respectively, expected in the static spin structure factor for the $\mathbb{Z}_2$ vortex crystal. Each color corresponds to a different spin-component as listed in panel a).
}
\label{fig:lattice}
\end{figure}

Adding a finite Kitaev coupling $J_K$ immediately results in a negative energy eigenvalue and, therefore, destabilises the 120$^\circ$ ground state. We can still diagonalize for the eigenenergies perturbatively in $J_K$. In lowest order and in the long-wavelength limit the zero modes do not hybridize, and we obtain for the tilting mode a dispersion relation that is given in the long-wavelength limit, $|\vect k| \ll |\vect Q|$, by
\begin{align}
\varepsilon^{\rm tilt}_{\bf Q+k} \approx J_H S^2 \frac{3}{4} {\bf k}^2 -
2J_K S^2 \sum_{\gamma=x,y,z} \vect k \cdot \vect a_\gamma  \sin(\vect Q  \cdot \vect a_\gamma) ( e_3^\gamma)^2 .
\end{align} 
It becomes maximally negative for a wavevector 
\begin{align}
&{\bf k}_{\rm inst} =
\frac{J_K}{J_H} \frac{4}{3} \sum_{\gamma=x,y,z} \vect a_\gamma \sin(\vect Q  \cdot \vect a_\gamma) ( e_3^\gamma)^2 
\\\nn&
= \frac{J_K}{J_H}  \Big(\frac{1}{\sqrt{3}}\Big[( e_3^y)^2+( e_3^z)^2-2 ( e_3^x)^2) \Big], ( e_3^z)^2-( e_3^y)^2\Big)^T,
\end{align} 
that can be expressed in terms of the normal $\vect{ e}_3$. In the special case where the spins of the 120$^\circ$ ordering are confined within the $x$-$y$ plane and $\vect{e}_3 = \vect{\hat z}$, this wavevector is just given by $\vect k_{\rm inst} = J_K/J_H (1/\sqrt{3},1)^T$. So it is the tilting Goldstone modes that trigger the instability of the 120$^\circ$ antiferromagnetic ordering in the presence of a finite Kitaev interaction $J_K$.


\subsection{Incommensurate antiferromagnet: $\mathbb{Z}_2$-vortex crystal}

Indeed, allowing for a slowly spatially varying orthogonal frame $\vect e_i(\vect r)$
one finds in the limit $|J_K| \ll J_H$ the effective energy functional $\mathcal{E} = \int d^2 \vect r\, \mathcal{L}$ with 
\begin{align} \label{Func}
\mathcal{L} =  \frac{3 J_H S^2}{4}  \sum_{\gamma=x,y,z}  e^-_\gamma(\vect r) \left[-\nabla^2 - 2 i q_K \vect a_\gamma \cdot  \nabla  \right] e^+_\gamma(\vect r),
\end{align}
where $\vect{e}^\pm =  (\vect{e}_1 \pm i \vect{e}_2)/\sqrt{2}$. The Kitaev interaction induces a coupling $q_K= 2 J_K/(\sqrt{3} J_H)$ to constant gauge fields given by the triangular lattice vectors $\vect a_\gamma$, that can be identified as Lifshitz invariants as previously pointed out in Ref.~\cite{rous2012}. The magnetization can thus minimize its energy by allowing for a spatial modulation of the SO(3) order parameter on large length scales proportional to $1/q_K \propto J_H/J_K$.

\subsubsection{Luttinger-Tisza approximation}

The character of this modulated classical ground state can be obtained by minimizing the Hamiltonian treating the orthonormal constraint, $\vect e_i \cdot\vect e_j = \delta_{ij}$, or equivalently, $\vect{\hat \Omega}^2 = 1$, within an improved  Luttinger-Tisza approximation \cite{LuttingerTisza}. The latter is a good approximation for large length scales $q_K |\vect r| \gg 1$, or, alternatively, for small momenta $|\vect q| \ll q_K$. Note that this latter limit does not commute with $J_K \to 0$, and, as a consequence, does not smoothly connect with the Heisenberg point.

We start with the functional  
\begin{align}
&\mathcal{E} = \\
\nn&J_K S^2 \sum_{\braket{ij}} \vect{\hat \Omega}_i \cdot\vect{\hat \Omega}_j + J_K S^2 \sum_{\gamma\parallel\braket{ij}} \hat \Omega^\gamma_i \hat \Omega^\gamma_j 
- \sum_i \lambda_i (\vect{\hat \Omega}^2_i - 1).
\end{align}
The unit length of the vector $\vect{\hat \Omega}_i$ is locally imposed with the help of the Lagrange multipliers $\lambda_i$.
Upon spatial Fourier transformation, $\vect{\hat\Omega}(\vect r)=\sum_{\vect q}e^{i\vect{qr}}\vect{\hat\Omega}_\vect{q}$, the functional takes the form 
\begin{align} \label{LuttTiszafunctional}
\mathcal{E}/N = \sum_{\vect{q}} \hat \Omega^\alpha_{-\vect q} \mathcal{J}^{\alpha \beta}(\vect q) \hat \Omega^\beta_{\vect q}
+ \sum_{\vect{q},\vect{p}} \lambda_{- \vect q - \vect p} \hat \Omega^\alpha_{\vect p} \hat \Omega^\alpha_{\vect q}
- \lambda_0,
\end{align}
where $\lambda_0 = \lambda_{\vect q}|_{\vect q=0}$. The matrix $J^{\alpha \beta}$ possesses only diagonal entries with 
\begin{align}
\mathcal{J}^{\alpha\alpha}(\vect q)
&= J_H S^2 \Bigl(\cos(\vect a_x\cdot \vect q )+\cos(\vect a_y \cdot\vect q )+\cos(\vect a_z \cdot \vect q )\Bigr) 
\nn\\&+ J_K S^2 \cos(\vect a_\alpha \cdot \vect q )
\end{align}
and $J^{\alpha \beta}(\vect q) = 0$ for $\alpha \neq \beta$. At the Heisenberg point $J_K = 0$, the diagonal components of the matrix are minimal for momenta at the corner of the Brillouin zone, e.g.\ $\vect q = \vect Q$, thus leading to 120$^\circ$ ordering. A finite $J_K$, however, favours in general incommensurate order with wave vectors away from $\vect Q$ as $J^{\alpha \alpha}(\vect q)$ become minimal for momenta of the form $\vect q^{(1)}_\alpha = \vect Q - t\, \vect a_\alpha$ with $t \in \mathbb{R}$. On the other hand, Fourier components, $\hat \Omega^\alpha_{\vect q^{(1)}_{\alpha}}$, of the spin with such incommensurate wave vectors induce finite Fourier components, $\lambda_{\pm 2 \vect q^{(1)}_\alpha}$ with $\alpha = 1,2,3$, of the Lagrange multiplier. Finite Lagrange multipliers $\lambda_{\pm 2 \vect q^{(1)}_\alpha}$, in turn induce two finite secondary Fourier components $\hat \Omega^\alpha_{\vect q^{(2)}_{\alpha,\beta}}$ with $\vect q^{(2)}_{\alpha, \beta} = \vect Q - t\, (2\vect a_\beta-\vect a_\alpha)$ where $\beta \neq \alpha$ and so on.

In the following, we discuss a Luttinger-Tisza approximation where we limit ourselves to the lowest finite Fourier components $\hat \Omega^\alpha_{\vect q^{(1)}_{\alpha}}$ and $\hat \Omega^\alpha_{\vect q^{(2)}_{\alpha,\beta}}$ for the spin and $\lambda_0$ and $\lambda_{\pm 2 \vect q^{(1)}_\alpha}$ for the Lagrange multiplier; all higher Fourier modes are neglected. In principle, this approximation can be systematically improved by including higher order modes.
Minimizing the functional \eqref{LuttTiszafunctional} within this approximation we obtain for the energy per site 
\begin{align} \label{LTEnergy}
\varepsilon_{\rm LT}(t) = -\frac{S^2}{9} &\Big[ J_H
\Big( \cos\frac{\pi + 6 t}{3} + 17 \sin\frac{\pi- 3 t}{6} 
\\\nn&+ 8 \sin\frac{\pi+ 6 t}{6} + \sin\frac{\pi + 15 t}{6} \Big)
\\\nn&
+  J_K \Big( \cos\frac{\pi + 6 t}{3} + 8 \sin\frac{\pi+ 6 t}{6} \Big)\Big],
\end{align}
which still depends on the parameter $t$ that quantifies the distance of the primary Bragg peak from the corner of the Brillouin zone, $\vect q^{(1)}_\alpha = \vect Q - t\, \vect a_\alpha$. The value of $t_{\rm min}$ identifying the position of the minimum of the function \eqref{LTEnergy} finally determines the ground state energy $\varepsilon_{\rm LT}(t_{\rm min})$. This analytical estimate for the ground state energy is found to be in excellent agreement with numerical estimates obtained from Monte Carlo simulations discussed in Sec.~\ref{sec:energies}.

The corresponding state is given by 
\begin{align} \label{VortexGS}
&S^\gamma(\vect r) \approx \frac{4 S}{3\sqrt{3}} {\rm Re} \Big\{ e^{i \phi}   \times
\\ \nonumber
&\Big( e^{i (\vect Q - t \vect a_\gamma)\cdot (\vect r-\vect r_0)} 
+ \frac{1}{4} \sum_{\eta \neq \gamma} e^{i (\vect Q - t (2 \vect a_\eta - \vect a_\gamma)) \cdot(\vect r-\vect r_0)} \Big)\Big\},
\end{align}
where $S^\gamma$ is the $\gamma$-component of the spin and the ground state is obtained by setting $t = t_{\rm min}$. The first term in Eq.~\eqref{VortexGS} is the most important, primary Fourier component which also possesses the  smallest deviation of momentum from the corner of the Brillouin zone, $\vect Q$. The secondary Fourier components have a smaller weight and are shifted further away by $- t_{\rm min} (2 \vect a_\eta - \vect a_\gamma)$ with $\eta \neq \gamma$.
The resulting Bragg peaks in the static structure factor are visualized in Fig.~\ref{fig:lattice}~b), which nicely agrees with previous numerical findings for the classical model \cite{rous2012}. The relative weight of secondary and primary Bragg peaks are predicted to be $1/4^2 = 1/16$ within the above approximation.
We find that the corresponding energy is independent of the choice of origin $\vect r_0 = (x_0,y_0)^T$ as well as the phase $\phi$.

In the Luttinger-Tisza approximation the length of the $\vect{\hat \Omega}$ vector is compromised to differ from unity, $\sum^3_{\alpha=1}(\hat \Omega_t^\alpha(\vect r))^2 \neq 1$. Whereas the length $|\vect{\hat \Omega}_t(\vect r)|$ varies in space it nevertheless  remains always finite so that the orientation of $\vect{\hat \Omega}_t(\vect r)$ is always well defined. Note that in the limit $J_K \to 0$ the distance $t_{\rm min}\to 0$ and $\varepsilon_{\rm LT}(0) = - S^2 3J_H/2$ recovers the exact ground state energy whereas the state itself, $\hat \Omega_{t=0}^\alpha(\vect r)$, does not reproduce the 120$^\circ$ ordering as expected.
 
\subsubsection{Vector chirality and $\mathbb{Z}_2$ vortices}

\begin{figure}[t]
\centering
\includegraphics[width=\linewidth]{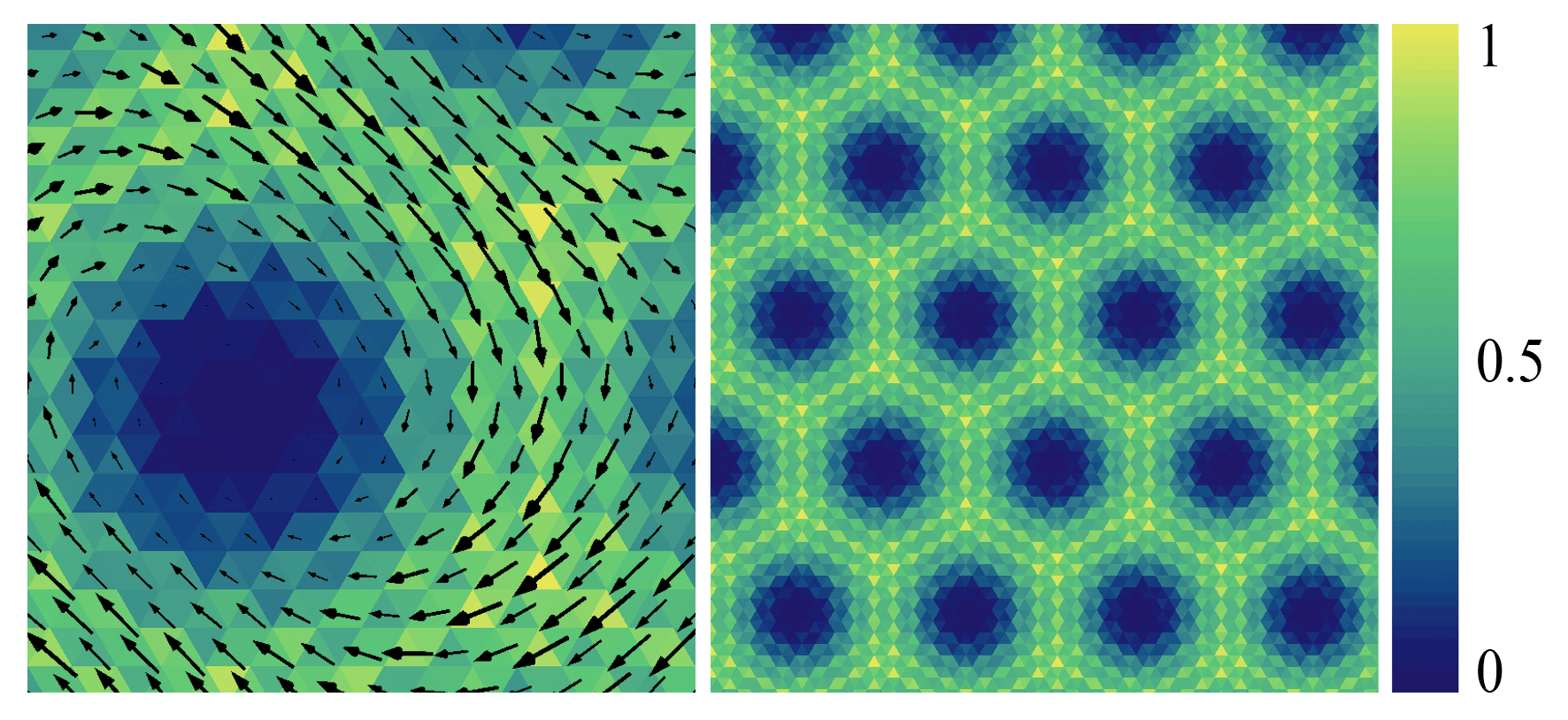}
\caption{(Color online)  $\mathbb{Z}_2$ vortex crystal stabilized for $J_H>0$ in the presence of a small but finite Kitaev interaction $J_K$ revealed by the chirality vectors of Eq.~\eqref{ChiralityVector} which were computed from the classical ground state \eqref{VortexGS} in the Luttinger-Tisza approximation. The colour code shows the length of the chirality vector, $|\vect{\kappa}(\vect{r})|$, normalized to one, that becomes minimal at the $\mathbb{Z}_2$ vortex cores. The arrows in the close-up of the left panel correspond to projections of $\vect{\kappa}(\vect{r})$ onto the $x$-$y$ plane.}
\label{fig:chirality}
\end{figure}

It turns out that the approximate classical ground state \eqref{VortexGS} corresponds to a triangular lattice of condensed $\mathbb{Z}_2$ vortices, thus confirming the numerical results of Ref.~\onlinecite{rous2012}.
This is best seen  by defining chirality vectors on upward pointing triangles of the lattice
\begin{align} \label{ChiralityVector}
\vect{\kappa}(\vect{r})=\tfrac{2}{3\sqrt{3}}\left(\vect{S}_{\vect{r}}\times\vect{S}_{\vect{r}+\vect{a}_x}+\vect{S}_{\vect{r}+\vect{a}_x}\times\vect{S}_{\vect{r}+\vect{a}_y}+\vect{S}_{\vect{r}+\vect{a}_y}\times\vect{S}_{\vect{r}}\right).
\end{align}
The length of $\vect{\kappa}(\vect{r})$ measures the rigidity of the local 120$^\circ$ ordering and it vanishes 
at the center of each $\mathbb{Z}_2$ vortex \cite{kawamura84}. The chirality vector profile, that derives from Eq.~\eqref{VortexGS}, is shown in Fig.~\ref{fig:chirality} and clearly reveals the $\mathbb{Z}_2$ vortex crystal.


Within our Luttinger-Tisza approximation we find three zero modes for the $\mathbb{Z}_2$ vortex crystal represented by  the phase $\phi$ and the vector $\vect r_0$. The latter are expected as the vortex crystal spontaneously breaks translational symmetry so that a constant shift of the origin $\vect r_0$ does not cost any energy. The corresponding low-energy excitations are just the effective acoustic phonon excitations of the vortex crystal. 
If the coupling between the two-dimensional atomic triangular lattice planes of Ba$_3$IrTi$_2$O$_9$ is sufficiently small, these low-energy modes will destroy true long-range order of the $\mathbb{Z}_2$ vortex crystal at any finite temperature that will be reflected in a characteristic broadening of the Bragg peaks in the structure factor of Fig.~\ref{fig:lattice}~b).

\subsection{Polarized neutron scattering}
\label{Sec:Neutrons}

The structure factor of the $\mathbb{Z}_2$ vortex crystal possesses as a hallmark of the Kitaev interaction a characteristic correlation between the positions of the Bragg peaks and the associated spin-components, see Fig.~\ref{fig:lattice}~b). We suggest to resolve this correlation with the help of spherical neutron polarimetry. 

The probability that an incoming neutron with spin $\vect \sigma_{\rm in}$ is scattered into a spin-state $\vect \sigma_{\rm out}$ is given by the energy-integrated scattering cross section $\sigma_{\vect \sigma_{\rm out},\vect \sigma_{\rm in}}(\vect q)$, where $\vect q$ is the transfered momentum. Consider a polarizer and analyzer with an orientation specified by the unit vectors $\vect e_{\rm in}$ and $\vect e_{\rm out}$, respectively. The total probability and the relative probability that a neutron is detected with spin $\pm \vect e_{\rm out}$ is then given by 
\begin{align}\vect
\sigma(\vect q, \vect e_{\rm out}, \vect e_{\rm in}) &= \sum_{\tau_{\rm out} = \pm 1} 
\sigma_{\tau_{\rm out}\vect \sigma_{\rm out},\vect \sigma_{\rm in}}(\vect q),
\\
\Delta \sigma(\vect q, \vect e_{\rm out}, \vect e_{\rm in}) &= \sum_{\tau_{\rm out} = \pm 1} 
\tau_{\rm out} \sigma_{\tau_{\rm out}\vect \sigma_{\rm out},\vect \sigma_{\rm in}}(\vect q),
\end{align}
respectively. The polarization is then defined by the ratio $\mathds{P}(\vect q, \vect e_{\rm out}, \vect e_{\rm in}) = \Delta \sigma(\vect q, \vect e_{\rm out}, \vect e_{\rm in})/\sigma(\vect q, \vect e_{\rm out}, \vect e_{\rm in})$. In the following, we concentrate on the magnetically ordered phase when the scattering probabilities are dominated by magnetic Bragg scattering so that we can neglect all nuclear contributions. For the particular choice that the axis of polarizer and analyzer coincide, $\vect e_{\rm out} = \vect e_{\rm in} \equiv \vect e$, but are orthogonal to the transfered momentum $\vect e \perp \vect q$, the polarization attributed to magnetic scattering simplifies to \cite{Blume1963,Maleyev1962}
\begin{align}
\mathds{P}_{\rm mag}(\vect q, \vect e, \vect e)\Big|_{\vect{\hat e }\perp \vect q} = 2 \frac{e_i \chi_{ij}(\vect q) e_j}{\chi_{kl}(\vect q)(\delta_{kl} -  \hat{q}_k  \hat{q}_l)} - 1,
\end{align}
where $\vect{\hat q}=\frac{\vect{q}}{|\vect{q}|}$ is the orientation of momentum and $\chi_{ij}(\vect q) = \chi_{ij}(\vect q,\omega=0)$ is the spin susceptibility at zero frequency,
\begin{align}
\chi_{ij}( \vect q,\omega) = i \int_0^\infty \text{d}t\; e^{i \omega t} \langle [{\bf S}_{i}(\vect q, t),{\bf S}_j(- \vect q, 0)] \rangle.
\end{align}

The magnetic structure factor of the $\mathbb{Z}_2$ vortex crystal, that follows from Eq.~\eqref{VortexGS}, has only non-zero diagonal components, $\chi_{ii}$, which however differ from each other and, moreover, possess different Bragg peak positions. For example, for our choice of the Kitaev interaction the $\chi_{zz}$ component is expected to exhibit a primary Bragg peak at $\vect q^{(1)} = \vect Q - t \vect a_z = \frac{1}{a}(\frac{4\pi}{3} (1,0,0)- t (-\frac{1}{2}, - \frac{\sqrt{3}}{2}),0)$ where $a$ is the lattice constant and we assumed for simplicity that the two-dimensional triangular lattice lies in the $x$-$y$ plane. Measuring at this particular Bragg peak, one expects for $\vect e = \vect{\hat{ z}}$ the value $\mathds{P}_{\rm mag} = 1$ in contrast to $\mathds{P}_{\rm mag} = -1$ that is obtained for $\vect e$ in the direction perpendicular to $\vect{\hat z}$ and $\vect q$. A systematic variation of the analyzer/polarizer orientation $\vect e$ should therefore allow, in principle, to resolve the correlation between the diagonal components $\chi_{ii}$ and their Bragg peak position.

\section{Full Phase diagram}
\label{sec:PhaseDiagram}

\begin{figure}[t]
\centering
\includegraphics[width=\linewidth]{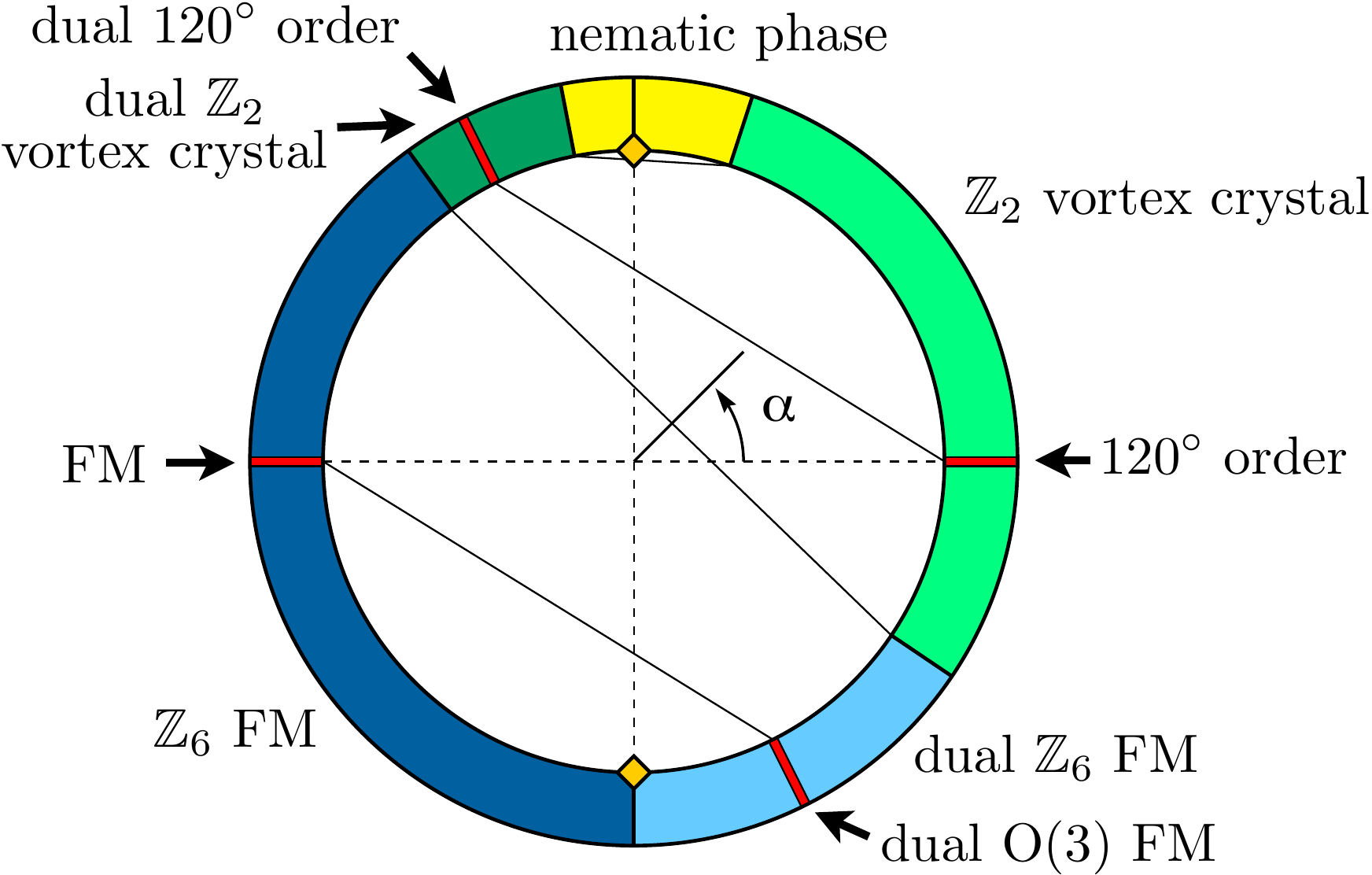}
\caption{(Color online) Phase diagram of the Hamiltonian \eqref{Ham} with parametrization $(J_H,J_K)=(\cos\alpha, \sin\alpha)$ as obtained from exact diagonalization data. Solid lines show the mapping between two Klein-dual points. Red lines mark the location of the four SU(2)-symmetric points. Yellow diamonds mark the two Kitaev points.}
\label{fig:phasediag}
\end{figure}

\begin{figure}[t]
\centering
\includegraphics[width=\linewidth]{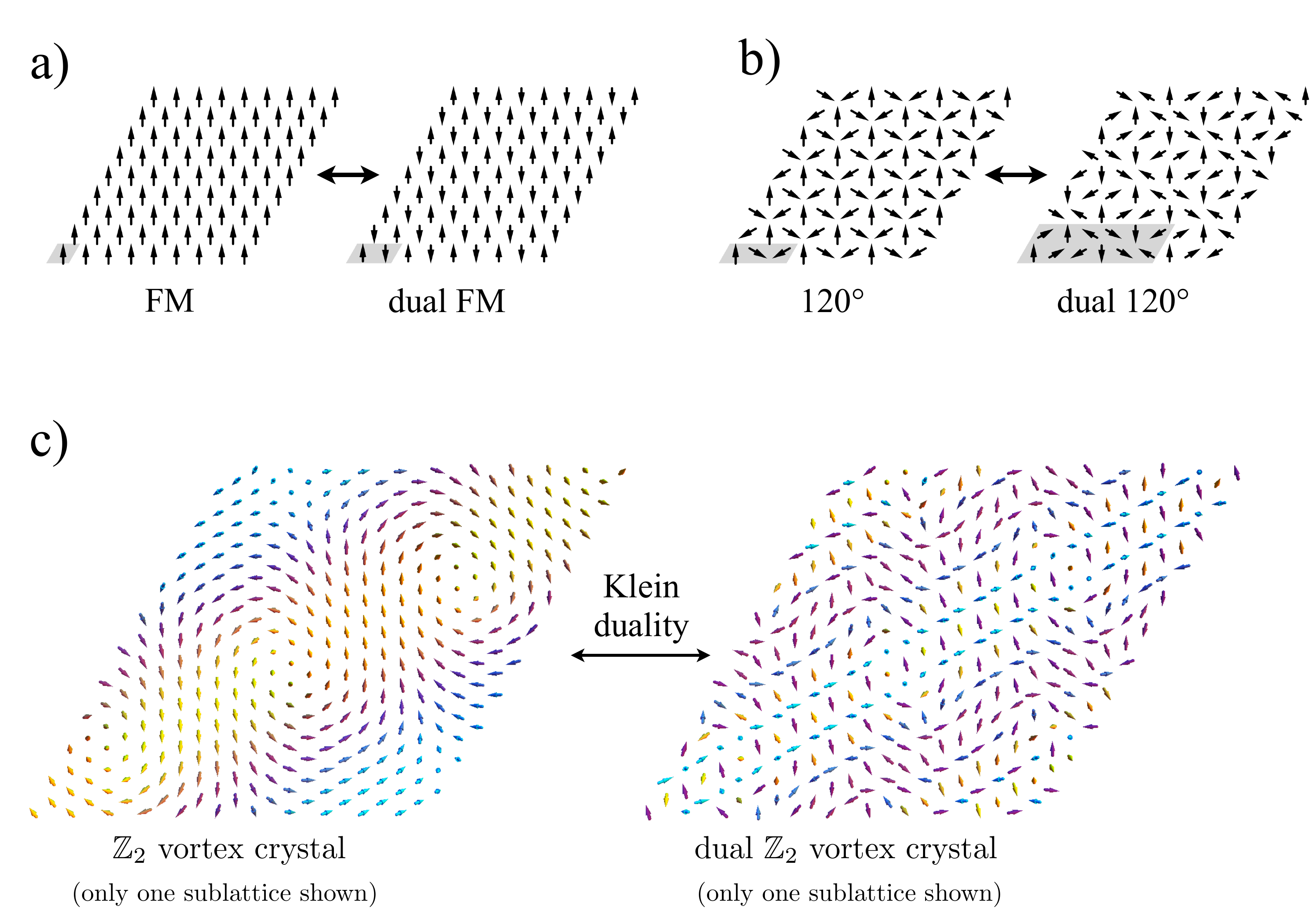}
\caption{(Color online) (a) and (b) Spin configurations for the four SU(2) symmetric points of the HK model \eqref{Ham}. The gray diamonds indicate the unit cells of the order.
(c) Snapshots of spin configurations in the $\mathbb{Z}_2$-vortex crystal (left) and its dual $\mathbb{Z}_2$-vortex crystal (right). For clarity, only one of the three sublattices of the triangular lattice is shown.Yellow arrows point upwards out of the plane, while blue arrows point downwards out of the plane. }
\label{fig:DualSpinConfigurations}	      
\end{figure}

After a detailed discussion of the magnetic structure close to the antiferromagnetic Heisenberg point in the previous section, we now turn to the remaining part of the phase diagram. It is represented in Fig.~\ref{fig:phasediag} by a circle with the help of the parametrization $(J_H,J_K)=(\cos\alpha, \sin\alpha)$. 

Importantly, the HK model \eqref{Ham} exhibits a duality \cite{Khaliullin2002,Khaliullin2005} (also referred to as the Klein duality \cite{Kimchi2014}) relating a pair of interactions on the right-hand side of the circle to a pair of interactions on the left-hand side, i.e. $J_H \to -J_H$ and $J_K \to 2J_H + J_K$. The corresponding dual states are related by a four-sublattice basis transformation, see appendix \ref{app:Duality} for more explanations. 
As a consequence, the antiferromagnetic, $\alpha = 0$, as well as the ferromagnetic Heisenberg point, $\alpha = \pi$, both possess a dual giving rise to four SU(2) symmetric points marked by red bars in Fig.~\ref{fig:phasediag}.  
In particular, this maps the ferromagnetic state for $J_H<0$ at $\alpha = \pi$ to a dual ferromagnet at $J_H>0$ and $J_K<0$ consisting of alternating strips of up and down pointing spins, see Fig.~\ref{fig:DualSpinConfigurations} a). 
Similarly, the 120$^\circ$ ordered state and its surrounding $\mathbb{Z}_2$-vortex crystal phase around the $J_H>0$ Heisenberg point map to a dual phase in the upper left quadrant with $J_H<0$ and $J_K>0$ with the respective orderings illustrated in Fig.~\ref{fig:DualSpinConfigurations} b) and c). 

In the following, we first elaborate in Sec. \ref{sec:FM} on the ferromagnetic phase and the influence of a finite Kitaev interaction on the order parameter space. Second, in Sec. \ref{sec:Kitaev} we examine the physics close to the Kitaev point $\alpha = \pi/2$ where the classical ground state manifold is macroscopically degenerate so that quantum fluctuations have a profound effect. Third, in Sec. \ref{sec:energies} we finally discuss the ground state energies of the classical as well as of the quantum model that lead to the phase diagram in Fig.~\ref{fig:phasediag}

\subsection{$\mathds{Z}_6$ ferromagnet}
\label{sec:FM}

At the Heisenberg point $J_H < 0$ and $J_K = 0$, the exact ground state of the Hamiltonian is the ferromagnetic spin-configuration where the order parameter is allowed to cover the whole sphere $S^2$, i.e., to point in any direction. In the presence of a finite $J_K$, however, fluctuations discriminate between the various orientations of the ferromagnetic order parameter and reduce the order parameter space from the sphere to $\mathbb{Z}_6$, i.e., to only six points. A similar order-by-disorder mechanism has recently been discussed~\cite{OBDO} with regard to distortions in the hexagonal HK model.

We concentrate here on the regime of the phase diagram adjacent to the ferromagnetic Heisenberg point (dark blue shaded in Fig.~\ref{fig:phasediag}). With the help of the duality transformation analogous conclusions then apply to the dual ferromagnet corresponding to the light blue shaded regime in Fig.~\ref{fig:phasediag}.

\subsubsection{Analytical arguments}

The classical ferromagnetic ground state is given by a constant, homogeneous spin configuration, $\vect{\hat \Omega}(\vect r) \equiv \vect{\hat \Omega}$ with $\vect{\hat \Omega}^2 = 1$. 
The corresponding classical energy per site is independent of the orientation of $\vect{\hat \Omega}$ and reads
\begin{align} \label{classicalFMenergy}
\varepsilon_{\rm FM} =  S^2 \Big(3 J_H + J_K \Big) 
\end{align} 
For $J_K = 0$, this indeed corresponds to the exact ground state energy. Any finite $J_K$, however, gives rise to fluctuation corrections to the ground state energy that also discriminate between the various orientations of $\vect{\hat \Omega}$. The leading $1/S$-fluctuation correction to the energy is computed in appendix \ref{app:FM} and reads in lowest order in the Kitaev interaction $J_K$
\begin{align}
\delta \varepsilon_{\rm FM} 
&= - \frac{S}{2} \frac{J_K^2}{|J_H|} 
\frac{3 (2\sqrt{3} - \pi)}{8\pi}
\Big(1 + \hat \Omega_x^4 + \hat \Omega_y^4+ \hat \Omega_z^4 \Big).
\label{eq:fmfit}
\end{align}
This correction favors the vector $\vect{\hat \Omega}$ to point along one of the six equivalent $\langle 100 \rangle$ directions (as $2\sqrt{3} - \pi > 0$). 
Whereas at the Heisenberg point, $J_K=0$, the ferromagnetic ground state manifold is the full sphere, $S^2$, a finite Kitaev interaction reduces this manifold to only six points corresponding to a $\mathds{Z}_6$ ferromagnetic order parameter.  

\subsubsection{Numerical evidence}
\begin{figure}
\centering
\includegraphics[width=\linewidth]{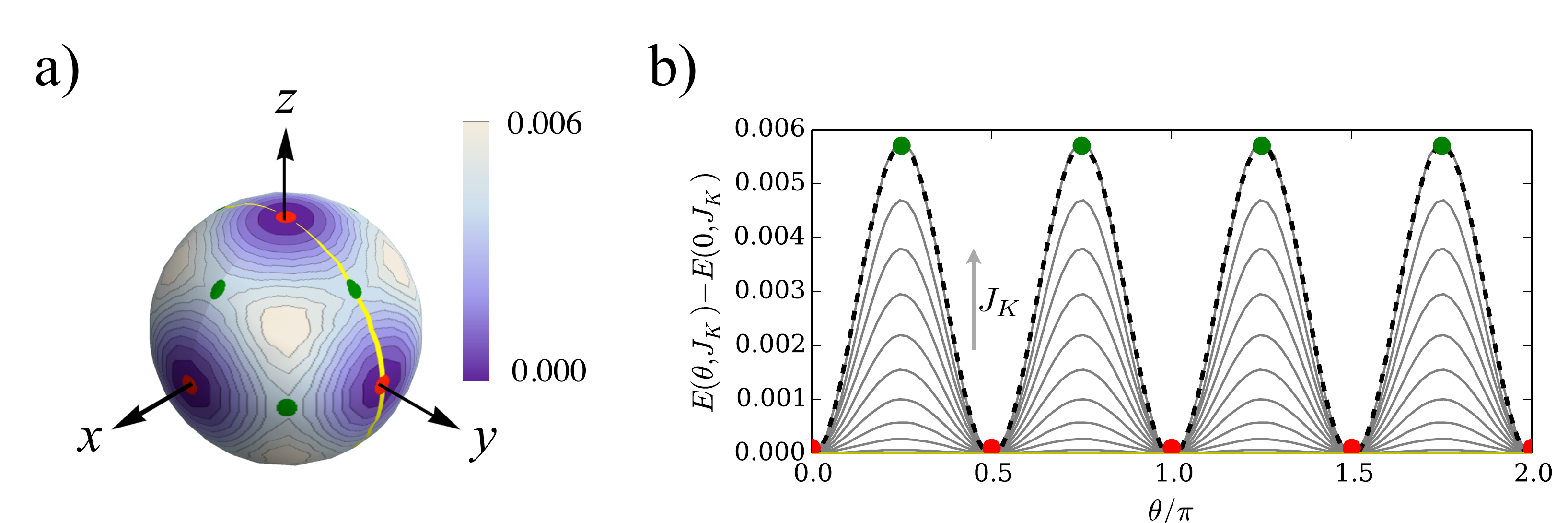}
\caption{(Color online) a) Ground state energy of the quantum model for ferromagnetic $J_H<0$ in an external Zeeman field as a function of the direction of the applied magnetic field $\vect{B}$, where we have subtracted the ground state energy for $\vect{B}=|\vect B| \vect{\hat{z}}$. The Kitaev coupling strength is $J_K/|J_H| = \tan(11\pi/10)\approx 0.32$. The energy is minimal when the magnetization is pinned along one of the three axes, and maximal when pointing along the space diagonals. b) The same results shown for the cut along the yellow line in a). Each line in b) corresponds to a different value of $J_K/J_H$. 
While for $J_K=0$ the ground state energy is independent of the direction of the magnetic field, the directional dependence becomes increasingly pronounced upon increasing $J_K/|J_H|$.
 The dashed line is a fit  of Eq.~\eqref{eq:fmfit}.}
\label{fig:EnergyCorrection}
\end{figure}

\begin{figure}
\centering
\includegraphics[width=\columnwidth]{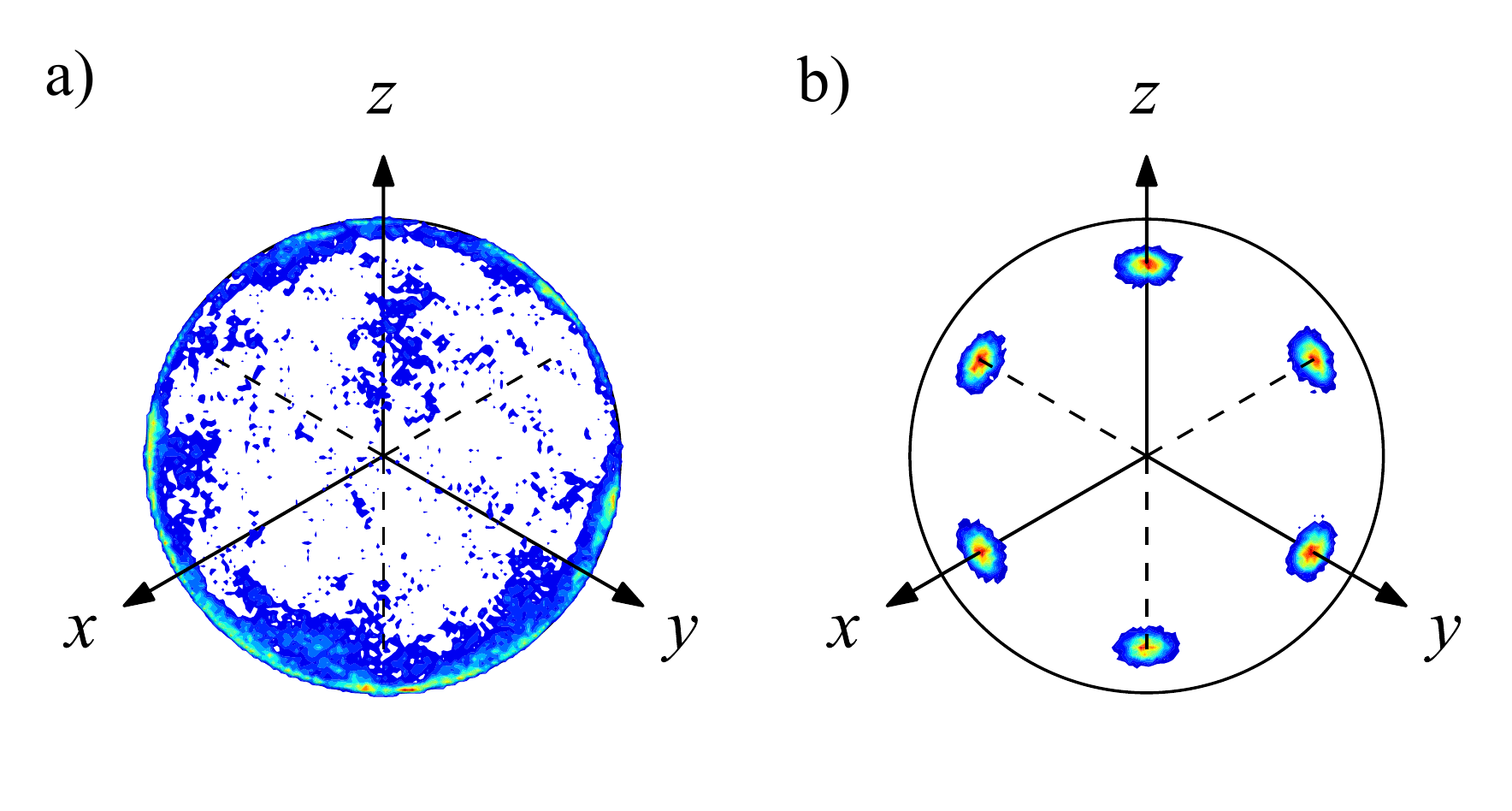}
\caption{(Color online) Histogram of the spin expectation value obtained with the help of finite-temperature Monte Carlo simulations  of the classical HK model close to the ferromagnetic Heisenberg point. Whereas for $J_K=0$ in panel (a) the spin covers the full $S^2$ sphere, {\em thermal} fluctuations in the presence of a finite $J_K \neq 0$ favor the alignment along one of the six $\langle 100\rangle$ directions. 
}
\label{fig:montecarlo}
\end{figure}

To corroborate our analytical results for the reduced order parameter space for $J_K\neq 0$ around the ferromagnetic Heisenberg point, we performed exact diagonalization calculations on small systems. We implemented lattice clusters with periodic boundary conditions containing 12 sites, with a geometry that preserves the $C_6$ rotational symmetry of the triangular lattice. By applying a small magnetic field $\vect{B}$ to each spin,
\begin{equation}
\vect{B} = B\begin{pmatrix}
\cos(\phi)\sin(\theta)\\
\sin(\phi)\sin(\theta)\\
\cos(\theta)
\end{pmatrix},
\end{equation}
where $\phi\in[0, 2\pi)$ and $\theta\in[0,\pi]$, the magnetization was forced to point in different directions. Fig.\ \ref{fig:EnergyCorrection} a) shows results for the change in the ground state energy as a function of the orientation of $\vect{B}$ with respect to the parallel alignment $\vect{B}\parallel\vect{\hat z}$ for a small finite Kitaev coupling $J_K/|J_H| = \tan(11\pi/10) \approx 0.32$. In agreement with our analysis above, the ground state energy of the system is minimal when the magnetization points along one of the six $\langle 100\rangle$ directions. Scanning the orientation of $\vect{B}$ along the yellow line shown in Fig.\ \ref{fig:EnergyCorrection} a), we compare in Fig.\ \ref{fig:EnergyCorrection} b) the effect of different Kitaev couplings (solid lines).
While for $J_K=0$ the energy is independent of the orientation of $\vect{B}$, for any finite $J_K\neq 0$ the energy immediately acquires an orientational dependence, that becomes more pronounced as $J_K$ increases. The black dashed line in Fig.\ \ref{fig:EnergyCorrection}b) is a fit of Eq.~(\ref{eq:fmfit}), showing perfect agreement.

The same reduction of the order parameter space is already at work on the classical level. Fig.~\ref{fig:montecarlo} shows result of a finite-temperature Monte Carlo simulation of the classical HK model. Whereas for $J_K=0$ the order parameter covers the $S^2$ sphere uniformly as illustrated in Fig.~\ref{fig:montecarlo} a), the thermal fluctuations in the presence of a finite $J_K$ favor the alignment of the order parameter along one of the six $\langle 100\rangle$ directions as shown in Fig.~\ref{fig:montecarlo} b).

\subsection{Nematic order close to the Kitaev point}
\label{sec:Kitaev}

\begin{figure}[t] 
\includegraphics[width=\linewidth]{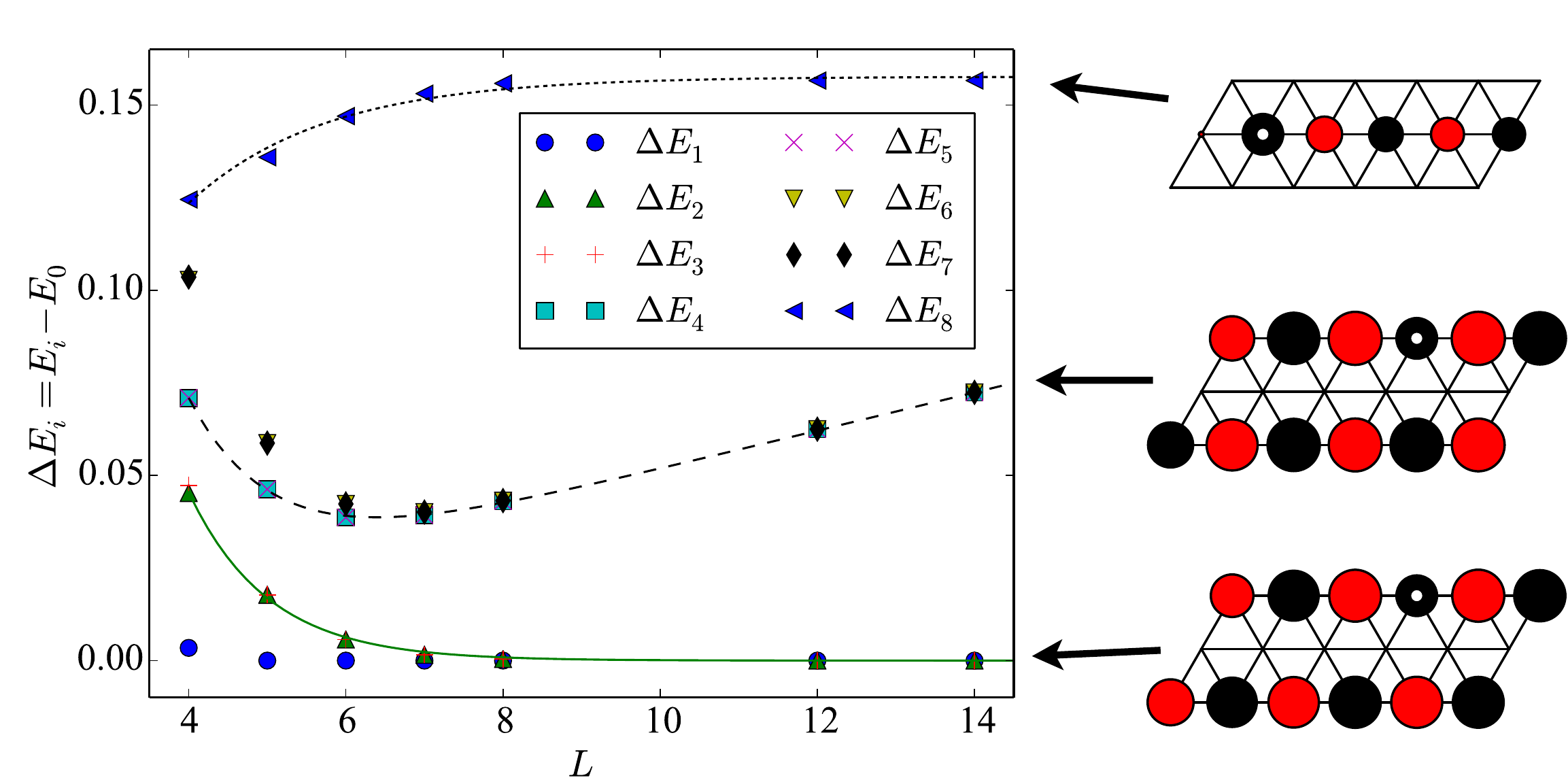}
\caption{(Color online) Energy gaps of a $3\times L$ triangular lattice strip with open boundary conditions. All values are given in relation to the ground state energy $E_0$, i.e.\ $\Delta E_1=E_1-E_0$. The figures on the right show numerical results for $\braket{S^x_{\vect{r}_0} S^x_{\vect{r}}}$ spin correlations, where the black disk with the white dot indicates the position $\vect{r}_0$, the diameter of the disks indicates the strength of the correlation and the color indicates the sign, with red corresponding to negative (antiferromagnetic) and black to positive (ferromagnetic) correlations. For details, see the main text.}
\label{fig:energies_3leg}
\end{figure}

In the classical limit, the Kitaev model on the triangular lattice possesses a macroscopic ground state degeneracy as pointed out in Ref.~\onlinecite{rous2012}. The spins form anti- or ferromagnetically ordered Ising chains, for $J_K > 0$ and $J_K < 0$, respectively, along one of the three lattice directions. The Kitaev interaction, however, does not couple the ordering of the individual chains thus giving rise to a $3\times 2^L$-fold sub-extensive ground state degeneracy where $L$ is the linear system size. Each ground state breaks the combined symmetry of the HK Hamiltonian of a C$_6$ lattice rotation and a cyclic spin exchange so that the ordering is that of a spin nematic.
While the ferromagnetic Kitaev point, $J_K < 0$, only separates the ferromagnetic and the dual ferromagnetic order, which is immediately stabilized for any finite $J_H$, an extended nematic phase arises close to the antiferromagnetic Kitaev point, $J_K > 0$ \cite{rous2012}. For later reference, the energy per site of the classical ground state close to the antiferromagnetic Kitaev point is given by
\begin{align}
\label{KitaevEnergy}
\varepsilon_{\rm nematic} = - S^2 (J_H + J_K).
\end{align}

In order to investigate this nematic ordering of the quantum model, we calculated the energies of the ground state and  the first few excited states using the density matrix renormalization group (DMRG) \cite{White,Dolfi2014}. Once the ground state was found, we targeted excited states by successively calculating states of lowest energy that are orthogonal to all previously found states. While the DMRG is highly successful for 1D systems, it can also be extended to systems with a small finite width, and we considered triangular lattice systems of width 3 and 4 and varying length with open boundary conditions. We ran calculations at bond dimensions $M=600, 800, 1000$ making sure that the energies converged.

\begin{figure}[t] 
\centering
\includegraphics[width=\linewidth]{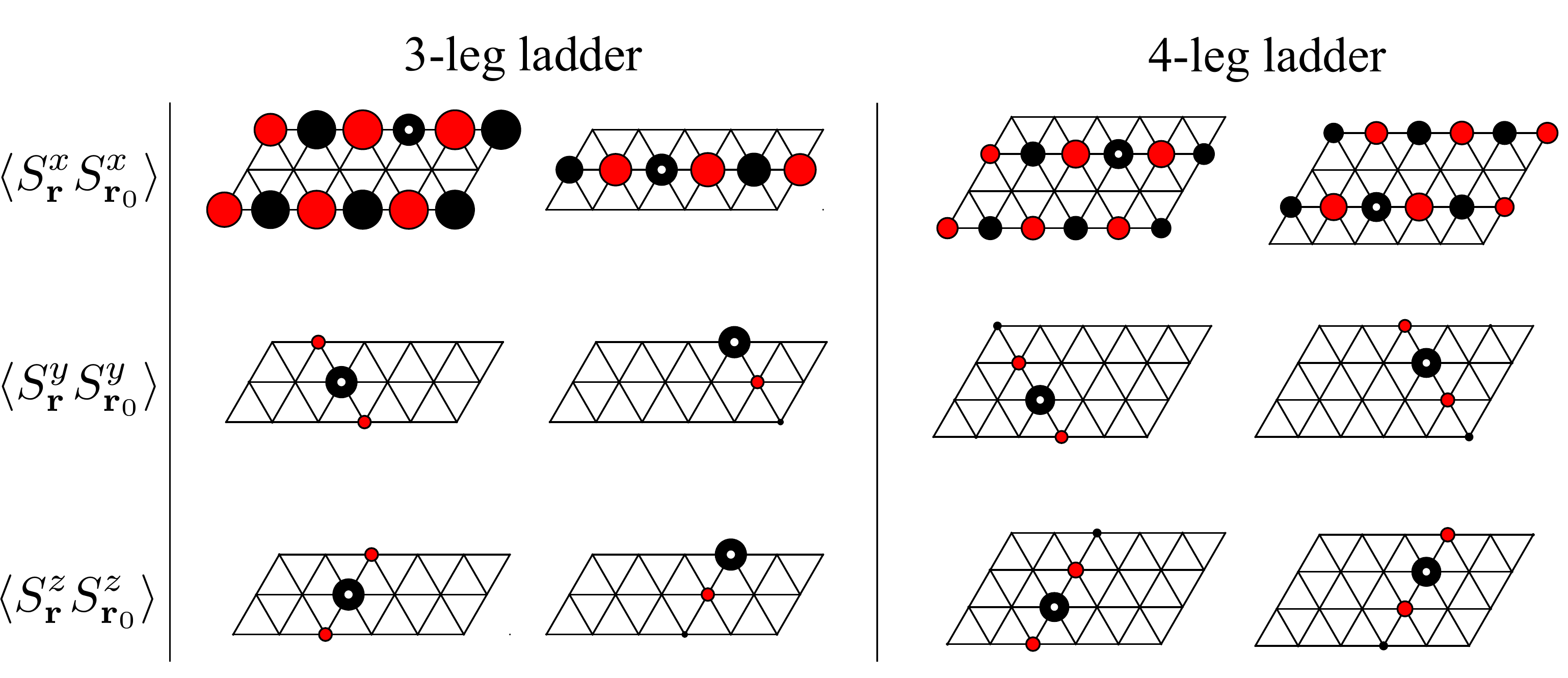}
\caption{(Color online) Spin-spin correlations in the ground state of the antiferromagnetic Kitaev model on the triangular lattice. Black circles indicate positive correlations, $\braket{S^\gamma_i S^\gamma_j} > 0$, whereas the red circles denote negative correlations. The small white dot indicates the position $\vect{r}_0$. The geometry of the lattice clusters lifts the degeneracy of the lattice direction, favoring 
chains antiferromagnetically coupled with their $x$-component along the $x$-direction while correlations along the $y$- and $z$-directions are suppressed. Whereas adjacent chains remain uncoupled, next-nearest neighbor chains couple antiferromagnetically.
}
\label{fig:kitaevcorrelators}
\end{figure}

The geometry of the considered lattice clusters breaks the $C_6$ symmetry of the lattice and the spins order antiferromagnetically in the spin component corresponding to the interaction term along the longer direction. In Fig.\ \ref{fig:energies_3leg} we show the energy differences between the lowest 8 excited states and the ground state, alongside spin-spin correlators. The first three excited states collapse exponentially onto the ground state energy as the length of the system increases. Likewise, the next four excited states collapse to the same energy, however growing linearly in system length. From the calculated spin-spin correlators we can identify this excitation to be given by a breaking of the antiferromagnetic ordering between next-nearest neighbor chains. Finally the 8th excited level corresponds to a local defect in a chain, which is indicated by the vanishing spin correlation in the center left corner of the lattice cluster. Fig.~\ref{fig:kitaevcorrelators} shows the spin-spin correlations in the ground states for systems of width 3 and 4 at the antiferromagnetic Kitaev point ($J_H=0$). While nearest neighbor chains are uncorrelated, there is a clear antiferromagnetic correlation between next-nearest neighbor chains in the spin component given by the chain direction. This mechanism locks the spin alignment of next-nearest neighbor chains to each other and thus reduces the macroscopic degeneracy of the ground state from $3 \times 2^L$ to the non-extensive value $3 \times 2^2$. Other spin components show only very short-ranged correlations as shown in the lower two panels of Fig.~\ref{fig:kitaevcorrelators}. Upon including a non-vanishing Heisenberg interaction correlations also form between nearest-neighbor chains further lifting the degeneracy to  $3\times 2$ states (not shown), which however preserve the nematic nature of the Kitaev point.

\begin{widetext}

\begin{figure}[t]
\includegraphics[width=\linewidth]{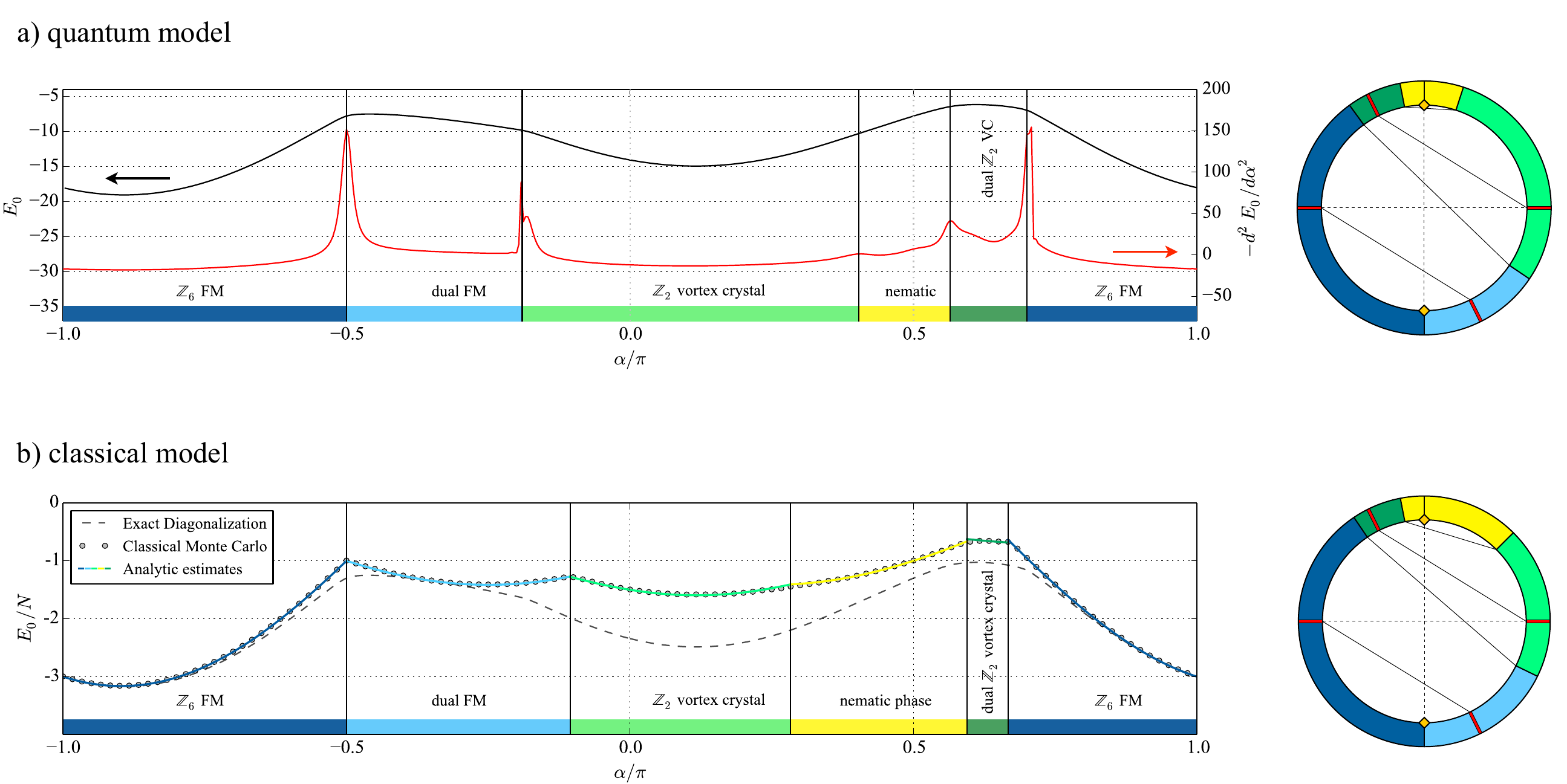}
\caption{
(Color online) {\it Upper panel (a):}
Ground state energy $E_0$ (black) and its second derivative, $-\text{d}^2E_0/\text{d}\alpha^2$, (red) for the HK quantum model obtained from exact diagonalization of small clusters. Peaks in the second derivative indicate the position of phase transitions. The black and red arrows indicate the corresponding axis for each data set.
{\it Lower panel (b):}
Classical energies (gray dots, Monte Carlo) and quantum energies (dashed, ED). 
The colored solid lines show analytical estimates for the classical ground state energies of the respective ordered phases, namely, Eq.~\eqref{classicalFMenergy} for the $\mathbb{Z}_6$ ferromagnet and its dual, Eq.~\eqref{LTEnergy} after minimization for the $\mathbb{Z}_2$-vortex crystal and its dual and Eq.~\eqref{KitaevEnergy} for the nematic phase.
The classical and quantum energies touch at the two fluctuation free points: the Heisenberg FM at $\alpha/\pi=1$ and its dual point.
The upper and lower ring summarize in the spirit of Fig.~\ref{fig:phasediag} the extension of the various phases of the quantum and the classical model, respectively.
}
\label{fig:Energies}
\end{figure}

\end{widetext}

\subsection{Phase boundaries and ground state energies}
\label{sec:energies}

The phase boundaries in Fig.~\ref{fig:phasediag} have been determined by calculating the ground state energy for clusters with $N=6\times4=24$ lattice sites and periodic boundary conditions as well as clusters with 27 lattice sites keeping the original $C_3$ lattice symmetry -- with both clusters  preserving the SU(2) symmetry of the Heisenberg points under the Klein duality.
Using exact diagonalization (ED) techniques, we have determined the phase boundaries by identifying the points where the second derivative $-\text{d}^2 E/\text{d}\alpha^2$ appears to diverge (on these finite systems), see the upper panel of Fig.~\ref{fig:Energies}.

For completeness, we have also repeated the Monte Carlo simulations of the classical model that were already performed in Ref.~\onlinecite{rous2012}. The result for the classical ground state energies is shown in the lower panel of Fig.~\ref{fig:Energies} together with a comparison to the ground state energies obtained from ED of the quantum model. As expected the two agree for the ferromagnetic Heisenberg model and its dual point indicating the absence of quantum fluctuations around their classical ground states. We also compare the Monte Carlo data with the analytical estimates for the classical ground state energies (colored solid lines), which approximate well the numerical result.
It should be noted that the phase diagram for the quantum HK model closely mimics the one found for the classical HK model \cite{rous2012}, which is due to the mainly classical nature of the various ordered phases. The exceptions are the Kitaev points where quantum fluctuations have a profound effect and lift the macroscopic degeneracy of the ground state.

\section{Conclusions}
\label{sec:Conclusions}

To summarize, we propose that a $\mathbb{Z}_2$-vortex crystal phase might be observed in the recently synthesized 
\Ba\  \cite{dey2012}. The latter forms a $j=1/2$ Mott insulator, whose low-energy physics we argue to be captured by a Heisenberg-Kitaev model on a triangular lattice. 
We reemphasize that the $\mathbb{Z}_2$-vortex crystal arises in the vicinity of the antiferromagnetic Heisenberg model, i.e. in the limit of small Kitaev interactions, and thus in the experimentally most relevant parameter  regime -- as revealed by numerous microscopic studies \cite{abinitio,models} of the honeycomb Iridates indicating the presence of Kitaev-type interactions only in addition to a dominant Heisenberg exchange.
Initial samples of \Ba\  \cite{dey2012} appear to suffer from significant Ir-Ti site inversion obscuring the formation of any ordered phase, but better samples should exhibit a distinct signature in polarized neutron scattering as we have discussed in detail.
The physics of the triangular HK model is also relevant to the honeycomb Iridates, for which it has been argued that a next-nearest neighbor exchange (along the two triangular sublattices of the honeycomb lattice) is indeed present in the actual materials \cite{Kargarian2012,reuther2012,Reuther14,Sizyuk14}.
Finally, we have left it to future research to explore whether the $\mathbb{Z}_2$-vortex crystal also plays out in the {\em bilayer} triangular lattice material Ba$_3$TiIr$_2$O$_9$ \cite{BaTiIrO}, which is closely related to the \Ba\  compound by replacing the role of Ir and Ti.

\acknowledgements
We acknowledge insightful discussions with M.~Daghofer, L.~Fritz, G.~Jackeli, M.~Punk, A.~Rosch, and I.~Rousochatzakis.
Our DMRG \cite{Dolfi2014} and exact diagonalization codes are based on the ALPS libraries~\cite{bauer2011-alps}. 
The numerical simulations were performed on the CHEOPS cluster at RRZK Cologne. Some of the figures were 
created using the Mayavi library \cite{mayavi} and Vesta \cite{vesta}, respectively.

\appendix

\section{Klein duality transformation}
\label{app:Duality}

\begin{figure}[b]
\centering
\includegraphics[width=\linewidth]{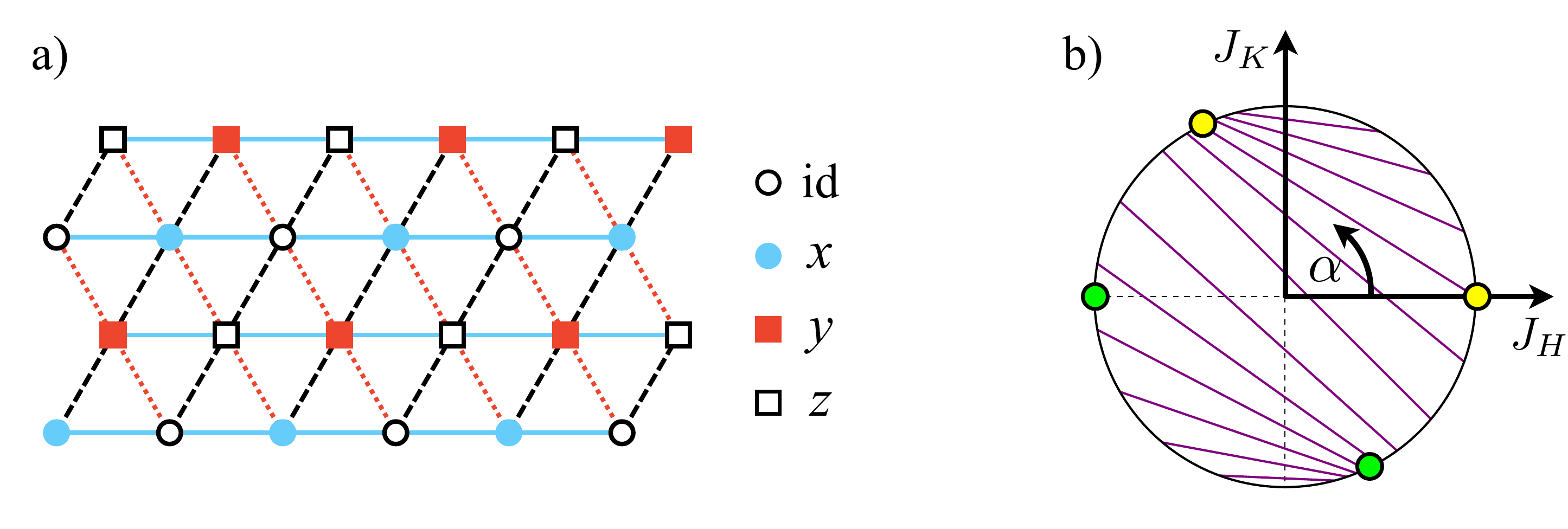}
\caption{(Color online)
	      a) 24 site cluster with periodic boundary conditions containing all symmetries except for the rotational $C_3$ symmetry.
		  The different symbols for the lattice sites indicate the four sublattices needed in the basis transformation underlying the Klein
		  duality \eqref{KleinDuality}.
	      b) Circle parametrization of the Heisenberg-Kitaev interactions $J_H = J \cos \alpha$ and $J_K = J \sin \alpha$ with the magenta
	      	  lines indicating points on the left and right-hand side of the circle related by the Klein duality \eqref{KleinDuality}.
		  The filled yellow and green circles indicate the points at which the Hamiltonian \eqref{Ham} is SU(2) symmetric.
	      }
\label{fig:cluster}
\end{figure}

We review the Klein duality relating couplings on the left and right-hand side of the circle phase diagram, see Fig.~\ref{fig:cluster}~b). Under this transformation, the Heisenberg-Kitaev Hamiltonian retains the same structure but the coupling parameters change as
\begin{align} \label{KleinDuality}
J_H \to -J_H,\qquad
J_K \to 2 J_H + J_K.
\end{align}
The transformation is performed by dividing the triangular lattice into four sublattices as illustrated in Fig.~\ref{fig:cluster}~a). Subsequently, each spin is subjected to a basis rotation, where the spins on the sublattice labeled ``id'' are not changed. For the three remaining sublattices each spin is rotated by $\pi$ around the spin axis according to the sublattice labeling. Since a $\pi$ rotation around one spin axis effectively inverses the sign of the two other components, we can write the full transformation as
\begin{subequations}
\begin{align}
&\text{id}: \qquad (S^x, S^y, S^z) \rightarrow (\phantom{-}S^x, \phantom{-}S^y, \phantom{-}S^z) \label{eq:klein_trafo_spins} \\ \nonumber
&x: \qquad (S^x, S^y, S^z) \rightarrow (\phantom{-}S^x, -S^y, -S^z) \\ \nonumber
&y: \qquad (S^x, S^y, S^z) \rightarrow (-S^x, \phantom{-}S^y, -S^z) \\ \nonumber
&z: \qquad (S^x, S^y, S^z) \rightarrow (-S^x, -S^y, \phantom{-}S^z).
\end{align}
\end{subequations}
Since this transformation is a simple local rotation of the spin basis, the original Hamiltonian and its counterpart after the transformation effectively describe the same physics, albeit for a resized unit cell. Interestingly, this transformation maps the SU(2) symmetric ferromagnetic and antiferromagnetic Hamiltonians at $J_K=0$ and $J_H=\pm1$ onto Heisenberg-Kitaev Hamiltonians with $J_K = -2 J_H$, revealing two more SU(2) symmetric points in the phase diagram. These points and their corresponding phases are termed the ``stripy'' (anti-)ferromagnets, due to the magnetic order after the basis rotation. The spin configurations at these points are illustrated in Fig.~\ref{fig:DualSpinConfigurations}.


\section{Fluctuation correction to the ferromagnetic ground state energy}
\label{app:FM}

The classical ferromagnetic ground state is given by a constant, homogeneous spin configuration, $\vect{\hat \Omega}(\vect r) \equiv \vect{\hat \Omega}$ with $\vect{\hat \Omega}^2 = 1$. 
The corresponding classical energy per site is independent of the orientation of $\hat \Omega$ and reads
\begin{align} \label{classicalFMenergy}
\varepsilon_{\rm FM} =  S^2 \Big(3 J_H + J_K \Big) 
\end{align} 
For $J_K = 0$, this indeed corresponds to the exact ground state energy. Any finite $J_K$, however, gives rise to fluctuation corrections to the ground state that also discriminate between the various orientations of $\vect{\hat \Omega}$. Performing a standard Holstein-Primakoff transformation, the spin-operator along the local $z$-axis, here defined by the classical vector $\vect{\hat \Omega}$, can be expressed as $\tilde {\vect S}_i^z = S - a^\dagger_i a_i$ where $a_i$ is a bosonic annihilation operator at the site $i$. Moreover,  
\begin{align}
\tilde {\vect S}_i^+ = \sqrt{2 S - a^\dagger_i a_i}\, a_i,\quad
\tilde {\vect S}_i^- = a^\dagger_i \sqrt{2 S - a^\dagger_i a_i}
\end{align}
where $\tilde {\vect S}_i^\pm = \tilde {\vect S}_i^x \pm i \tilde {\vect S}_i^y$. The spin-operator $\vect S$ within the laboratory frame is related to $\tilde {\vect S}$ by a rotation $\vect S = R \tilde{\vect S}$ where
\begin{align}
R = 
\left(\begin{array}{ccc} 
-\sin \phi & - \cos\theta \cos \phi & \sin\theta \cos\phi \\
\cos\phi & - \cos\theta \sin \phi &  \sin\theta \sin\phi \\
0 & \sin \theta & \cos \theta 
\end{array} \right).
\end{align}
and $\vect{\hat \Omega} = R(0,0,1)^T$.
Expanding the Hamiltonian in second order in the bosonic operators one obtains $\mathcal{H}= N \varepsilon_{\rm FM} + \mathcal{H}^{(2)}$ with 
\begin{widetext}
\begin{align}
\mathcal{H}^{(2)} &= \frac{1}{2} \sum_{\bf k \in {\rm 1. BZ}} (a^\dagger_{\vect k}\,a_{-\vect k}) h({\bf k}) \left(\begin{array}{c}
a_{\vect k} \\ a^\dagger_{-\vect k} \end{array}\right)
- \frac{S}{2} \sum_{\bf k \in {\rm 1. BZ}} \Big[
2 J_H  \sum_{\gamma = x,y,z} (\cos({\bf k}\cdot\vect a_\gamma) - 1) + 
J_K \sum_{\gamma = x,y,z} (\cos({\bf k}\cdot\vect a_\gamma) - 1) (1 - \hat \Omega^2_\gamma)
\Big] 
\end{align}
where  
\begin{align}
h({\bf k}) &= S \Big[
2 J_H  \sum_{\gamma = x,y,z} (\cos({\bf k}\cdot\vect a_\gamma) - 1) \mathds{1} + 2 J_K \Big\{
(\cos({\bf k}\cdot\vect a_x) - 1) 
\Big( e^+_x  e^-_x \mathds{1} + ( e^+_x)^2 \sigma^+ +  ( e^-_x)^2 \sigma^- \Big)
\\\nonumber
&\qquad +
(\cos({\bf k}\cdot\vect a_y) - 1) 
\Big(  e^+_y  e^-_y \mathds{1} + ( e^+_y)^2 \sigma^+ +  ( e^-_y)^2 \sigma^- \Big)
+
(\cos({\bf k}\cdot\vect a_z) - 1)   e^+_z  e^-_z
 (\mathds{1} - \sigma^x)
\Big\}\Big]
\end{align}
\end{widetext}
with the Pauli matrices $\sigma^x$,  $\sigma^y$, and $\sigma^z$, and we used the abbreviations $\vect{e}^\pm = \frac{1}{\sqrt{2}} R (1,\pm i,0)^T$ and $\sigma^\pm = \frac{1}{2}(\sigma^x \pm i \sigma^y)$. With the help of a Bogoliubov transformation we can compute the correction to the classical ground state energy \eqref{classicalFMenergy}. In order to elucidate the analytical structure, we concentrate on the contribution to this correction only of lowest order in the Kitaev interaction,
\begin{widetext}
\begin{align}
\delta \varepsilon_{\rm FM} &= - \frac{1}{4 N} \sum_{\bf k \in {\rm 1. BZ}} \frac{h_{21}({\bf k}) h_{12}({\bf k})}{h_{11}({\bf k})|_{J_K = 0}}
\\\nonumber
&= - \frac{S}{2 N} \frac{J_K^2}{|J_H|} \sum_{\bf k \in {\rm 1. BZ}} \frac{\big|
(\cos({\bf k}\cdot\vect a_x) - 1)( e^+_x)^2+(\cos({\bf k}\cdot\vect a_y) - 1)( e^+_y)^2 -
(\cos({\bf k}\cdot\vect a_z) - 1)   e^+_z  e^-_z\big|^2
}{\sum_{\gamma = x,y,z} (1- \cos({\bf k}\cdot\vect a_\gamma))}.
\end{align}
\end{widetext}
To evaluate this expression we need the following integrals over the Brillouin zone 
\begin{align}
&\frac{1}{N}\sum_{\bf k \in {\rm 1. BZ}} \frac{
(\cos({\bf k}\cdot\vect a_\alpha) - 1)(\cos({\bf k}\cdot\vect a_\beta) - 1)}{\sum_{\gamma = x,y,z} (1- \cos({\bf k}\cdot\vect a_\gamma))}  \\\nonumber
&\overset{N \to \infty}{\longrightarrow}
\frac{1}{\mathcal{V}_{\rm 1. BZ}} \int_{{\rm 1. BZ}} d {\bf k} \frac{
(\cos({\bf k}\cdot\vect a_\alpha) - 1)(\cos({\bf k}\cdot\vect a_\beta) - 1)}{\sum_{\gamma = x,y,z} (1- \cos({\bf k}\cdot\vect a_\gamma))}\\\nonumber
&= 
\frac{6 \sqrt{3}-2\pi}{3\pi} \delta_{\alpha \beta} + 
\frac{5\pi-6 \sqrt{3}}{6\pi}
(1-\delta_{\alpha \beta}).
\end{align}
Here, we evaluated the integrals in the thermodynamic limit where the volume of the first Brillouin zone is given by 
$\mathcal{V}_{\rm 1. BZ} = \frac{8\pi^2}{\sqrt{3}}$ using the identities
\begin{align}
&(  e^+_x   e^-_x)^2  + (  e^+_y  e^-_y)^2 + (  e^+_z   e^-_z)^2 =
\\
& -(  e^+_x   e^-_y)^2   
- (  e^+_y  e^-_x)^2 + ((  e^+_x)^2+(  e^+_y)^2)   e^+_z   e^-_z\nonumber\\&
+  e^+_z   e^-_z ((  e^-_x)^2+(  e^-_y)^2)\nonumber\\ \nonumber\
&=
\frac{1}{4}\Big(1 + \hat \Omega_x^4 + \hat \Omega_y^4+ \hat \Omega_z^4 \Big).
\end{align}
The fluctuation correction to the energy in lowest order in the Kitaev interaction finally assumes the form given in Eq.~\eqref{eq:fmfit}.
%


\bibliographystyle{apsrev4-1}

\end{document}